\documentclass{article}

\PassOptionsToPackage{numbers, compress}{natbib}
\usepackage{float}

    \usepackage[preprint]{neurips_2024}

\usepackage[utf8]{inputenc} 
\usepackage[T1]{fontenc}    
\usepackage{hyperref}       
\usepackage{url}            
\usepackage{booktabs}       
\usepackage{amsfonts}       
\usepackage{nicefrac}       
\usepackage{microtype}      
\usepackage{xcolor}         
\usepackage{svg}            
\usepackage{graphicx}
\usepackage{subcaption}
\usepackage{siunitx}        
\usepackage{multirow}

\title{Viability and Performance of a Private LLM Server for SMBs: A Benchmark Analysis of Qwen3-30B on Consumer-Grade Hardware}

\author{
  Alex Khalil \\
  UCLouvain\\
  \texttt{alex.khalil@student.uclouvain.be} \\
  \And
  Maria Parraga\\
  Universidad Espíritu Santo\\
  \texttt{mparraga@uees.edu.ec} \\
  \AND
   Guillaume Heilles\\
  DENEM Labs\\
  \texttt{guillaume@denemlabs.com} \\
  \And
  Simon Heilles\\
  DENEM Labs\\
  \texttt{simon@denemlabs.com} \\
}

\begin{document}

\maketitle

\begin{abstract}

Large Language Models (LLMs) are powerful AI systems but are usually only available through expensive cloud services owned by big tech companies. This creates problems for smaller organizations that need to protect their data, control their systems, and keep costs predictable. In this paper, we explore whether it is possible for small and medium-sized businesses (SMBs) to run a fast, capable LLM on their own hardware instead of relying on the cloud. We test an open-source model called Qwen3, optimized to fit on a high-end consumer GPU. We measure how well it performs on reasoning and knowledge tests, and how efficiently it runs when several users use it at once. Our results show that, with the right setup, a local LLM server can reach performance levels close to commercial cloud models, at a fraction of the cost and without giving up privacy.

\end{abstract}

\section{Introduction}
\label{sec:introduction}

The rapid development of Large Language Models (LLMs) has reshaped numerous industries, but access remains dominated by centralized, cloud-based services from large technology providers. While convenient, these services impose high and unpredictable costs and require sending sensitive data off-site, creating barriers for Small and Medium Businesses (SMBs).

This paper investigates the viability of a sovereign\footnote{By “sovereign,” we mean on-premise hosting of LLMs, under full organizational control, without reliance on third-party cloud providers.} LLM deployment that prioritizes both economic efficiency and data confidentiality. We hypothesize that increasingly powerful consumer hardware, combined with efficient quantized open-source models, provides a practical pathway to rival cloud performance. To test this, we build and evaluate a dedicated inference server based on the NVIDIA RTX 5090 running Qwen3-30B-A3B-Instruct, a Mixture-of-Experts (MoE) model \citep{shazeer2021switch}, with Q6\_K\_XL quantization.

Our benchmarking takes a dual approach: (1) evaluating the model’s intrinsic capabilities on standardized reasoning and knowledge benchmarks, and (2) stress-testing the server under varying concurrency to measure latency, throughput, and time-to-first-token. This design allows us to assess both model quality and system scalability under realistic SMB workloads.

\section{Background and related work}
\label{sec:background}

Our study lies at the intersection of three threads that shape the central question of this paper: quantization methods that make large models feasible on commodity GPUs; the motivations and constraints of sovereign (on-premise) LLM deployments; and systems and benchmarking work that capture inference behavior under realistic multi-user workloads.

\subsection{Quantization for consumer hardware}
\label{sec:quantization}

{\sloppy\hbadness=10000
Quantization, reducing weight and/or activation precision, is the main enabler for fitting large models on a single consumer GPU. Post-training weight-only and mixed-precision schemes commonly achieve $4{\times}$–$8{\times}$ memory reduction with minimal task degradation. Representative methods include GPTQ, a second-order, one-shot weight quantizer achieving 3–4-bit compression with negligible perplexity increase \citep{frantar2022gptq}, and activation-aware\slash block-wise schemes such as AWQ that preserve salient weights and enable hardware-friendly packing for 4–6-bit deployment \citep{lin2024awq}. Sparse-plus-quantized approaches (e.g., SpQR) further improve the compression–accuracy frontier by combining pruning with low-bit quantization \citep{dettmers2023spqr}.
\par}

{\sloppy\hbadness=10000
In practice, mid-bit formats (approximately 5–6 bits; e.g., Q6 variants) often provide the best latency–quality trade-off on 24–48\,GB GPUs lacking native ultra-low-bit support. Tooling for such formats (e.g., Q6\_K) is widely available in local stacks (llama.cpp, vLLM \citep{vllm2023}) and recent open model releases \citep{xia2024fp6,llamacpp2025}. Two implications guide our use of Q6\_\allowbreak K\_\allowbreak XL on an RTX 5090: mid-bit quantization comfortably fits 30B-class models (weights + KV cache) within single-GPU memory, and wall-clock gains depend on kernel\slash runtime support, benefits vanish if dequantization\slash packing dominates compute \citep{frantar2022gptq,lin2024awq}.
\par}

\subsection{On-Device and sovereign deployment}
\label{sec:sovereign}

{\sloppy
On-premise (sovereign) hosting is driven by confidentiality, regulatory compliance (e.g., GDPR\slash HIPAA), and predictable cost for sustained workloads. Policy and industry analyses argue that enterprises and public institutions increasingly prioritize sovereign models to retain control over data, deployment, auditability, and total cost of ownership (TCO) \citep{paloniemi2025onpremise,sovereignAI2024}. While cloud APIs minimize friction, usage-based billing can be volatile at scale; conversely, on-premise incurs upfront capital expense but can be cheaper at steady utilization while avoiding offsite data transfer.
\par}

Sovereign deployments, however, pose co-design challenges: firstly selecting minimal hardware that sustains acceptable multi-user QoE, secondly choosing quantization/serving pipelines that preserve model quality, and lastly operating an inference stack with controlled tail latency and energy. Reviews of on-device LLMs emphasize that these are joint model–runtime–hardware problems \citep{wang2023ondevicellms}. We therefore ask: today (2025) can a minimal server built around a consumer-grade GPU (RTX 5090) serve a sovereign 30B-class model with competitive QoE relative to online services?

\subsection{Benchmarks and comparative studies}

We measure two axes: \emph{model capability} and \emph{server operation}.

\textbf{Model quality}\\ Standard public suites (e.g. MMLU, AIME) remain the norm for assessing reasoning, breadth, and generality, enabling direct comparison between a locally hosted open model and proprietary services. The Qwen3 family (including Qwen3-30B-A3B) provides model cards and quantized artifacts that facilitate such evaluation \citep{qwen32025release}.

\textbf{System behavior}\\ Recent serving work identifies KV-cache management, batching, and scheduling as primary determiners of throughput and tail latency. PagedAttention\footnote{KV-cache optimizations like Paged Attention reduce memory overhead during long-context inference by dynamically managing attention states.} and the vLLM stack show that memory-aware scheduling and continuous/dynamic batching can deliver $2$–$4{\times}$ throughput improvements versus naïve pipelines, especially at long contexts \citep{kwon2023pagedattention}. Queueing-based approaches (e.g., FastServe) demonstrate that preemptive, length-aware admission control substantially reduces P95/P99 latency under mixed workloads \citep{zhang2023fastserve}. Comprehensive inference suites (e.g., LLM-Inference-Bench) motivate the metrics we adopt: end-to-end latency, time-to-first-token (TTFT), tokens/s, request throughput (RPS), and tail latency (P95/P99) \citep{chitty2024llminferencebench}. Kernel-level advances (FlashAttention family) remain central as attention/FFN memory I/O dominates latency and energy at long sequence lengths \citep{dao2022flashattention,shen2024flashattention2}.

\section{Methodology}

\label{Methodology}
\label{sec:methodology}
Our methodology is designed to evaluate whether a single consumer-grade GPU, the RTX 5090, can support a sovereign large language model (LLM) with multiple concurrent users while maintaining performance comparable to established cloud-based services. We describe the hardware and software environment, the model configuration, the evaluation benchmarks, the workload generation, and the measurement methodology.

\subsection{Hardware setup}
\begin{itemize}
    \item \textbf{GPU:} Nvidia RTX 5090 (32 GB, 3090 MHz boost clock, driver 580.65.06).
    \item \textbf{CPU:} 12th Gen Intel(R) Core(TM) i3-12100F (4 cores, 8 threads, up to 4.3 GHz).
    \item \textbf{System Memory:} Crucial 16 GB, DDR4, 3200 MT/s
    \item \textbf{Storage:} Crucial E100 Gen4 500 GB NVMe SSD (approx. 1316 MB/s read)
    \item \textbf{Power Supply:} BeQuiet Pure Power 12 1000W 80PLUS 
\end{itemize}

\subsection{Software environment}
\begin{itemize}
    \item \textbf{Operating System:} Linux Mint 22.1.
    \item \textbf{CUDA / cuDNN / Driver Versions:} 13.0/9.10.2/580.65.06.
    \item \textbf{Framework:} llama.cpp (commit 29c8fbe4e05fd23c44950d0958299e25fbeabc5c)
    \item \textbf{Tokenizer:} Byte-Pair Encoding (Qwen3Tokenizer).
\end{itemize}

\subsection{Model configuration}

\begin{itemize}
    \item \textbf{Model Size:} 30B parameters.
    \item \textbf{Quantization Scheme:} Q6\_K\_XL  (an effective 6.57 bits-per-weight scheme using 6-bit weights with high-precision scaling factors applied to blocks of 256)
    \item \textbf{Serving Parameters:} Max sequence length: 40960, max new tokens: 32768, temperature: 0.6, top-p: 0.95.
    \item \textbf{Batching and Scheduling Policy:} The server was configured with a continuous batching (-cb) policy, capable of processing up to 16 requests in parallel (--parallel 16).
\end{itemize}

\subsection{Evaluation benchmarks}
\paragraph{Quality Benchmarks}
\begin{itemize} 
    \item AIME 2024 \citep{aime2024_hf} (mathematical reasoning).
    \item AIME 2025 (mathematical reasoning).
    \item MMLU (broad and multidisciplinary knowledge assessment).
\end{itemize}
\paragraph{System benchmarks}
\begin{itemize}
    \item \textbf{Number of Concurrent Requests}: number of parallel user sessions sustained by the server.
    \item \textbf{Mean Input Tokens}, \textbf{Mean Output Tokens}, \textbf{Total Tokens Mean}: 
    distributional measures of sequence length.
    \item \textbf{Input Tokens per Request}, \textbf{Output Tokens per Request}, \textbf{Input–Output Ratio}: 
    structural features capturing request composition.
    \item \textbf{Tokens per Second (TPS)}: aggregate throughput across all requests.
    \item \textbf{Request Output Throughput Mean}, \textbf{Throughput per Concurrency}, 
    \textbf{Throughput per Pressure}: efficiency-normalized throughput indicators.
    \item \textbf{Completion Rate (RPS)}: completed requests per second, a proxy for sustained system capacity.
    \item \textbf{Token Demand per Second}: instantaneous load pressure from concurrent token generation.
    \item \textbf{End-to-End Latency Mean (E2E)}: mean latency from prompt submission to full response.
    \item \textbf{Time to First Token (TTFT)}: interactivity metric reflecting first-token delay.
    \item \textbf{Latency per Output Token}, \textbf{Latency per Total Token}: fine-grained latency breakdowns.
    \item \textbf{TTFT Fraction of E2E}: proportion of total latency attributable to first-token delay.
\end{itemize}

\subsection{Workload generation}
To evaluate multi-user performance, we generate synthetic request workloads using llmperf \citep{llmperf2024}.

\begin{itemize}
    \item \textbf{Concurrent Requests:} Varied between 1 and 16 to simulate a growing number of parallel users.  
    \item \textbf{Input Lengths:} Mean input tokens sampled between 16 and 2048, with standard deviation from 8 to 512, 
    to capture variability in prompt size.  
    \item \textbf{Output Lengths:} Mean output tokens ranged from 16 to 8192, with standard deviation from 4 to 512, 
    reflecting short and long generations.  
    \item \textbf{Request Arrival Rate:} Varied from 0.1 to 30 requests per second (req/s) to stress-test the system under 
    low and high demand.  
    \item \textbf{Completion Limits:} Maximum completed requests per run ranged between 50 and 200, ensuring sufficient 
    samples for statistical analysis.  
    \item \textbf{Sampling Parameters:} Temperature was fixed at 0.7, while Top-p varied between 0.2 and 1.0 to capture 
    different sampling strategies.  
    \item \textbf{Timeouts:} Each request was capped at a fixed timeout of 300 seconds to prevent indefinite hangs.  
\end{itemize}

\subsection{Baselines}
We compare against:
\begin{itemize}
    \item OpenAI ChatGPT o3
    \item OpenAI ChatGPT o1
    \item Google Gemini 2.5 Flash
    \item Anthropic Claude 3.7 Sonnet 
\end{itemize}

\section{Results}
\label{sec:results}

\subsection{LLM performance benchmarks}
\label{sec:llm_benchmarks}

We evaluate the locally hosted, quantized \texttt{Qwen3-30B-A3B} model on two mathematical reasoning datasets (AIME 2024 and AIME 2025) and on MMLU, capturing both high-precision multi-step reasoning and broad multidisciplinary knowledge transfer. All evaluations follow the harness and dataset sources described in the main text (LocalAIME, HuggingFace AIME2025, and the lm-evaluation-harness), using the model in its Q6\_K quantized deployment. Exact-match scoring is applied where defined, and no external tools (calculators, solvers) are used. This setup reflects the intended sovereign inference scenario on a single RTX 5090 GPU. The objective is not exhaustive leaderboard comparison but characterization of practical strengths and failure modes under realistic serving constraints. 

\subsubsection{AIME}

The AIME benchmark tests symbolic and numerical reasoning under strict correctness requirements. We report results for AIME 2024 and AIME 2025 using the public splits.

\begin{figure}[H]
  \centering
  \begin{subfigure}{0.48\linewidth}
      \centering
      \includegraphics[width=0.8\linewidth]{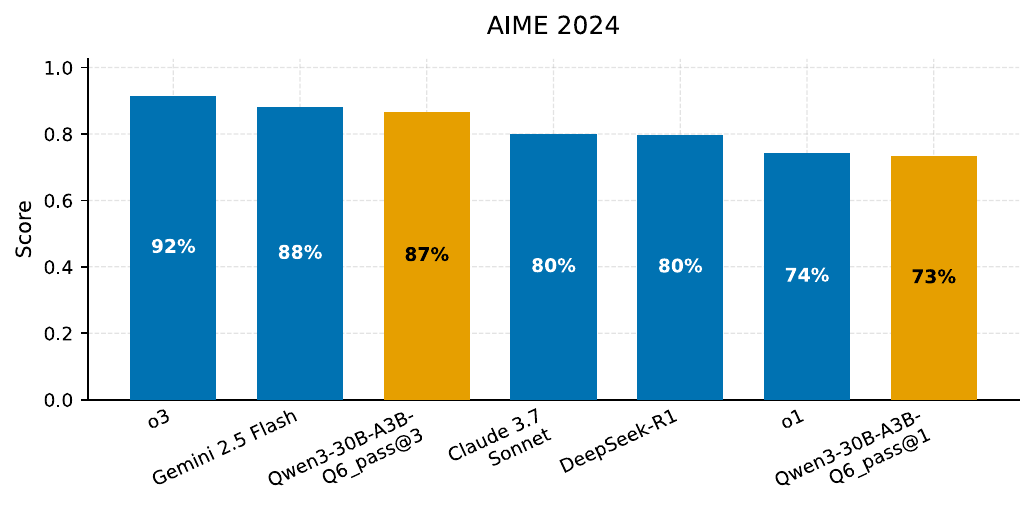}
      \caption{AIME 2024 — model accuracy compared with cloud-based baselines}
      \label{fig:aime2024}
  \end{subfigure}
  \hfill 
  \begin{subfigure}{0.48\linewidth}
      \centering
      \includegraphics[width=0.8\linewidth]{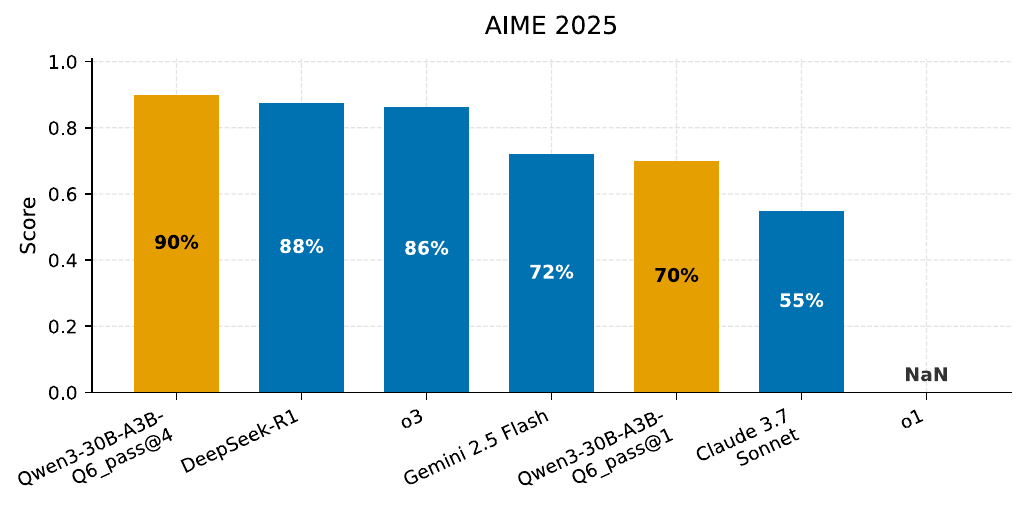}
      \caption{AIME 2025 — model accuracy compared with cloud-based baselines}
      \label{fig:aime25}
  \end{subfigure}
  \caption{Overall comparison of AIME model accuracy for 2024 and 2025.}
  \label{fig:aime}
\end{figure}

\noindent\textbf{Comment (AIME 2024)\footnote{Scores taken from \url{https://llm-stats.com/benchmarks/aime-2024}}:}  
The locally hosted Qwen3-30B-A3B (Q6\_K) achieves 73\%--87\% pass accuracy depending on sampling strategy, remaining competitive with strong proprietary systems (e.g., Claude 3.7 Sonnet and DeepSeek-R1 at 80\%).

\noindent\textbf{Comment (AIME 2025)\footnote{Scores taken from \url{https://llm-stats.com/benchmarks/aime-2025}}:} 
On the newer 2025 split, the model attains 70\%--90\% depending on sampling, outperforming Claude 3.7 Sonnet (55\%) but trailing o3 (86\%).

\subsubsection{MMLU}

MMLU evaluates general knowledge transfer across 57 subjects. We apply the multiple-choice scoring protocol from the lm-evaluation-harness \citep{lmeval2023} to enable comparability with community baselines.

\begin{figure}[h]
  \centering
  \includegraphics[width=0.4\linewidth]{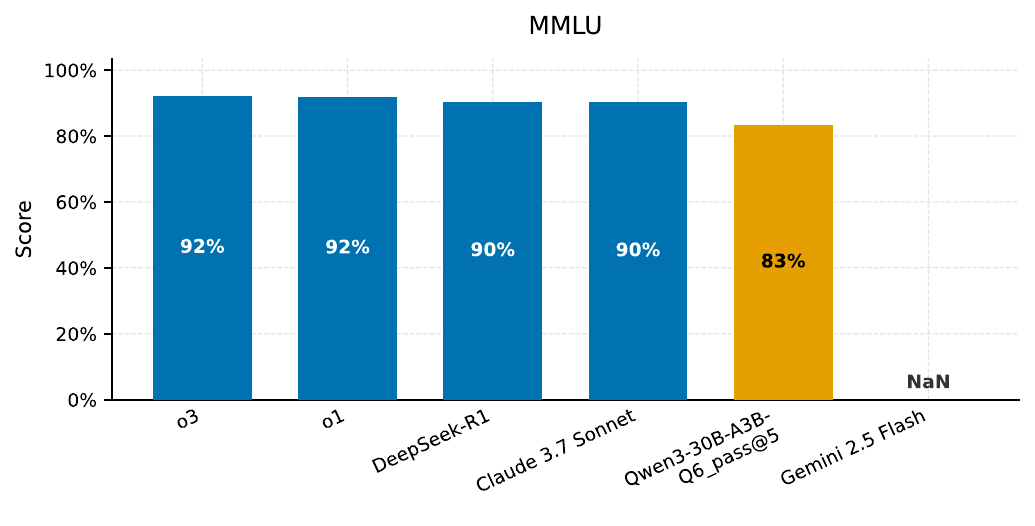}
  \caption{MMLU — model accuracy compared with cloud-based baselines}
  \label{fig:mmlu}
\end{figure}

\noindent\textbf{Comment (MMLU)\footnote{Scores taken from \url{https://www.kaggle.com/benchmarks/open-benchmarks/mmlu}}:}  
The Qwen3-30B-A3B (Q6\_K) reaches 83\% accuracy, placing slightly below frontier systems such as o3 and DeepSeek-R1 (92\%) but still within the high-performance band. Performance parity across diverse subjects demonstrates that quantization on a single RTX 5090 preserves broad factual knowledge, with degradation mainly on fine-grained or ambiguous items rather than structural reasoning.

\subsection{Server performance benchmarks}
\label{sec:server_benchmarks}

The benchmarking analysis was conducted on a server provisioned with an Nvidia RTX 5090 GPU, hosting the Qwen3-30B-A3B\_Q6 model. To ensure a comprehensive and statistically robust evaluation of the server's performance, we utilized Latin Hypercube Sampling (LHS) \citep{mckay1979lhs}.

\subsubsection{Spearman correlation matrix}
\label{sec:spearman}

\begin{figure}[htbp]
    \centering
    \includegraphics[width=0.5\textwidth]{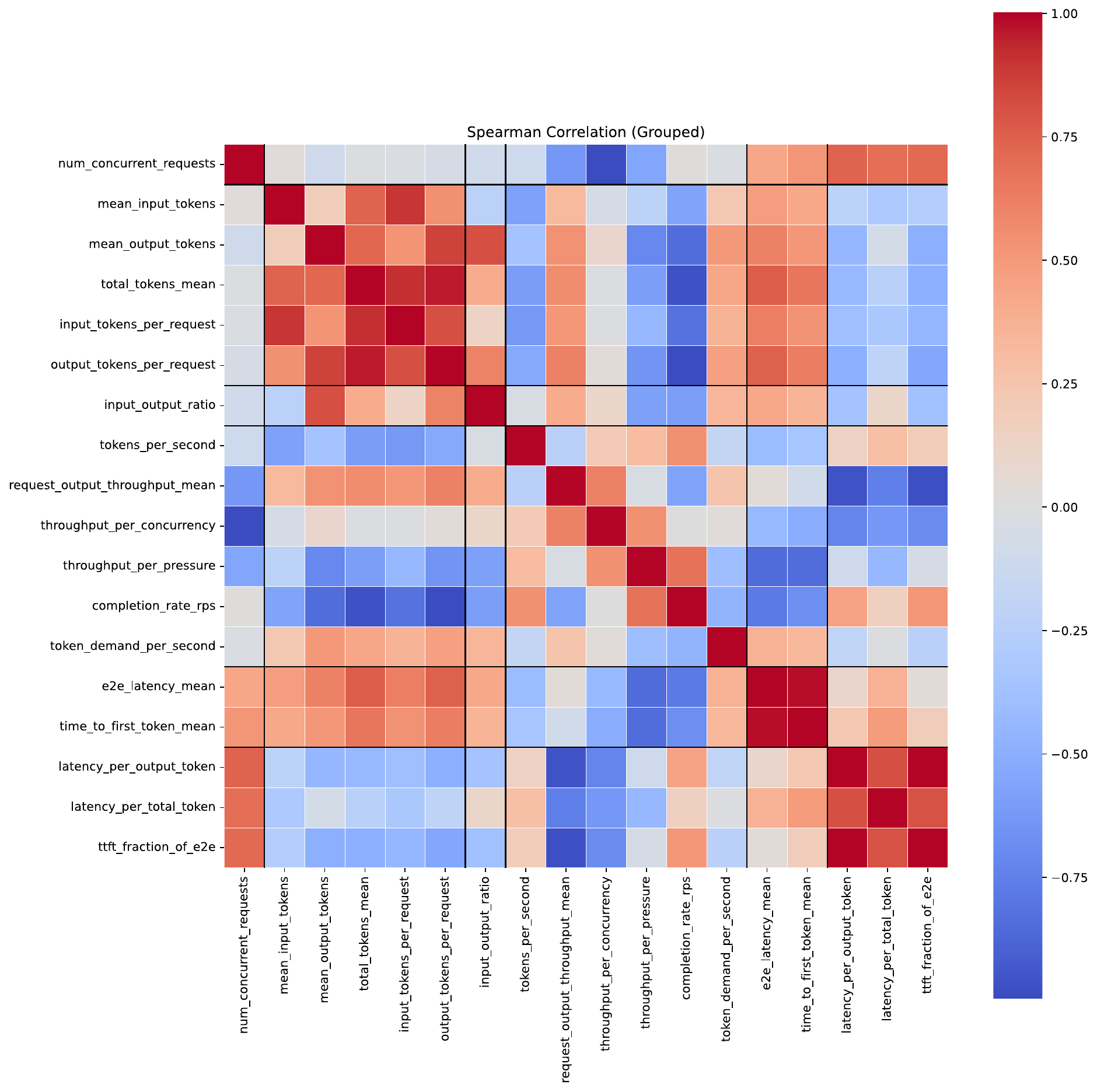}
        \label{fig:spearman}
    \caption{Spearman Correlation Matrix}
\end{figure}

{\sloppy
The Spearman correlation analysis reveals four coherent clusters: \texttt{sequence size} (mean\_\allowbreak input\_\allowbreak tokens, mean\_\allowbreak output\_\allowbreak tokens, total\_\allowbreak tokens\_\allowbreak mean, input\_\allowbreak tokens\_\allowbreak per\_\allowbreak request, output\_\allowbreak tokens\_\allowbreak per\_\allowbreak request, input\_\allowbreak output\_\allowbreak ratio), \texttt{efficiency} (tokens\_\allowbreak per\_\allowbreak second, request\_\allowbreak output\_\allowbreak throughput\_\allowbreak mean, throughput\_\allowbreak per\_\allowbreak concurrency, throughput\_\allowbreak per\_\allowbreak pressure, completion\_\allowbreak rate\_\allowbreak rps), \texttt{demand} (token\_\allowbreak demand\_\allowbreak per\_\allowbreak second, num\_\allowbreak concurrent\_\allowbreak requests), and \texttt{latency} (e2e\_\allowbreak latency\_\allowbreak mean, time\_\allowbreak to\_\allowbreak first\_\allowbreak token\_\allowbreak mean, latency\_\allowbreak per\_\allowbreak output\_\allowbreak token, latency\_\allowbreak per\_\allowbreak total\_\allowbreak token, ttf\_\allowbreak fraction\_\allowbreak of\_\allowbreak e2e).
\par}

Token counts, especially input size, are strongly positively correlated with all latency measures and negatively with efficiency, confirming that prompt processing dominates performance degradation. Output size correlates with overall latency but only weakly with per-token decode cost, underscoring that prefill, not generation, drives marginal slowdown. Latency metrics are highly collinear (e.g., latency\_per\_output\_token vs latency\_per\_total\_token), suggesting redundancy for modeling. The time to first token scales with input length but its fraction of total latency decreases with longer outputs, indicating amortization across decoding. Concurrency correlates negatively with throughput and positively with latency, showing queueing overhead rather than scalable gains. Throughput normalizations (per concurrency, per pressure) align tightly with tokens\_per\_second, while token\_demand\_per\_second bridges demand and latency, serving as an early pressure signal before collapse. 

Overall, the correlation structure confirms that \textbf{prompt footprint and concurrency are the dominant drivers of latency inflation and throughput erosion}, while output length has secondary impact. Thus, optimization should target prefill efficiency (compression, caching, flash/paged attention) and concurrency management (admission control, batching) to maintain near-linear efficiency under realistic multi-user workloads. 

\subsubsection{Performance analysis}
\label{sec:performance_analysis}
\subsection*{Impact of token count on latency and throughput}

\begin{figure}[htbp]
    \centering
    \begin{subfigure}{0.48\textwidth}
        \includegraphics[width=\linewidth]{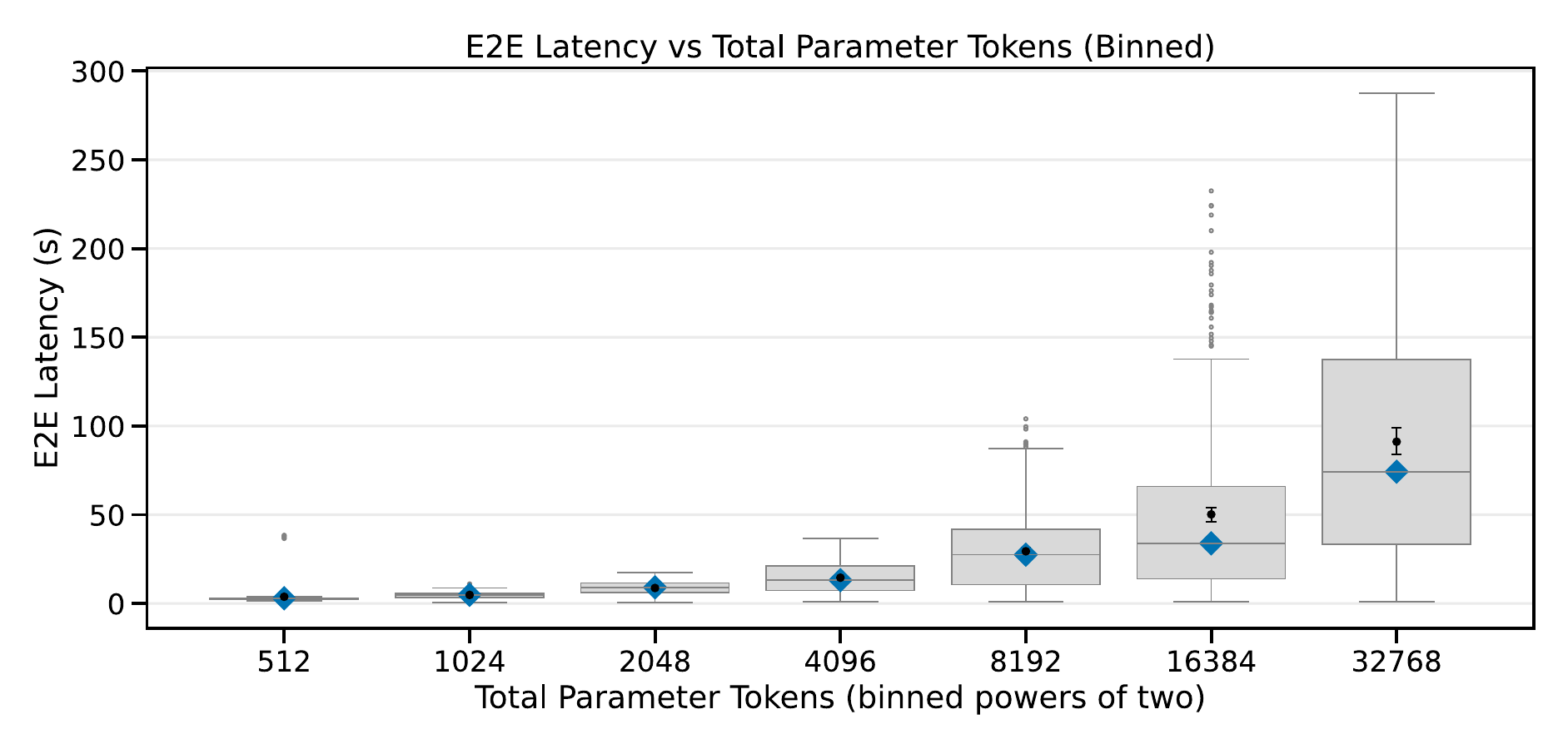}
        \caption{E2E Latency vs. Total\_Tokens (Binned)}
        \label{fig:latency_vs_tokens_binned}
    \end{subfigure}
    \hfill
    \begin{subfigure}{0.48\textwidth}
        \includegraphics[width=\linewidth]{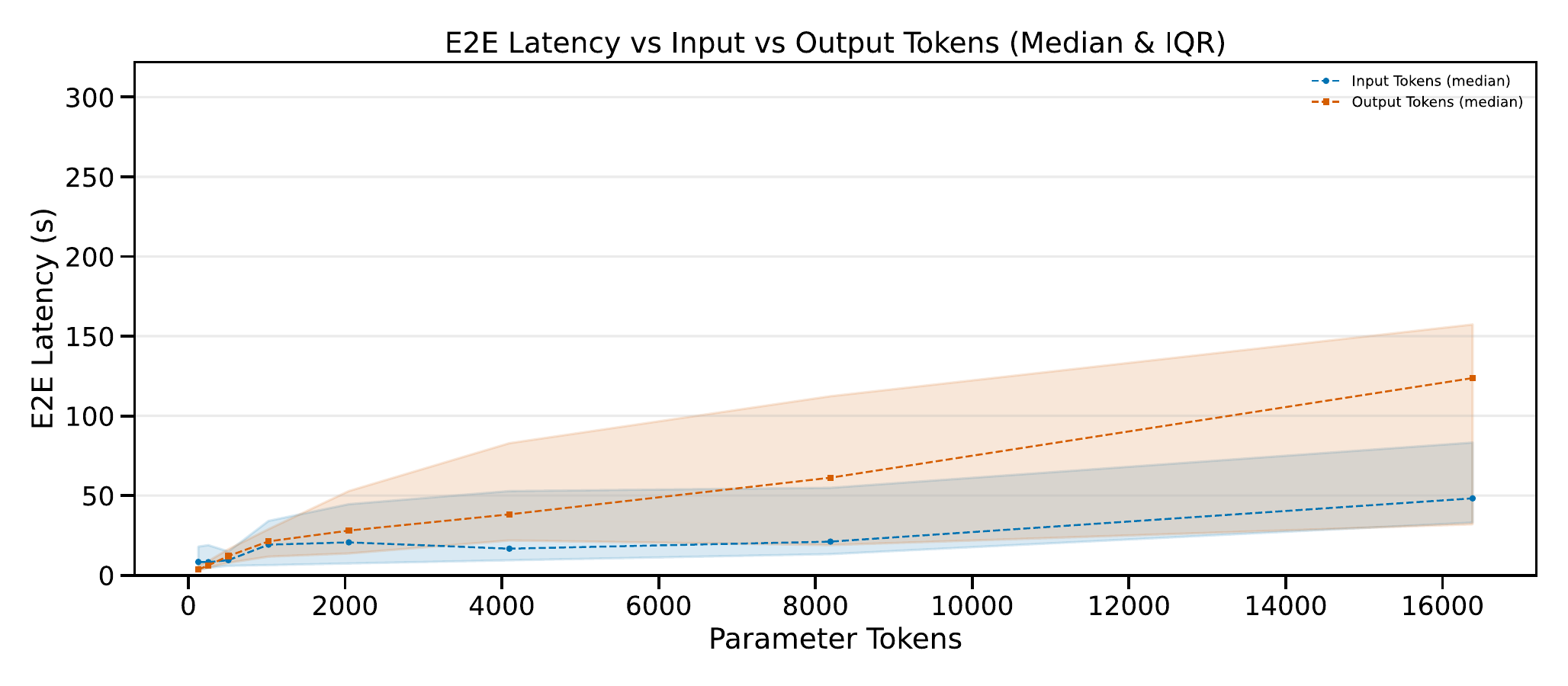}
        \caption{E2E Latency vs. Input/Output Tokens}
        \label{fig:latency_vs_tokens_line}
    \end{subfigure}
    
    \vspace{1em} 
    
    \begin{subfigure}{0.48\textwidth}
        \includegraphics[width=\linewidth]{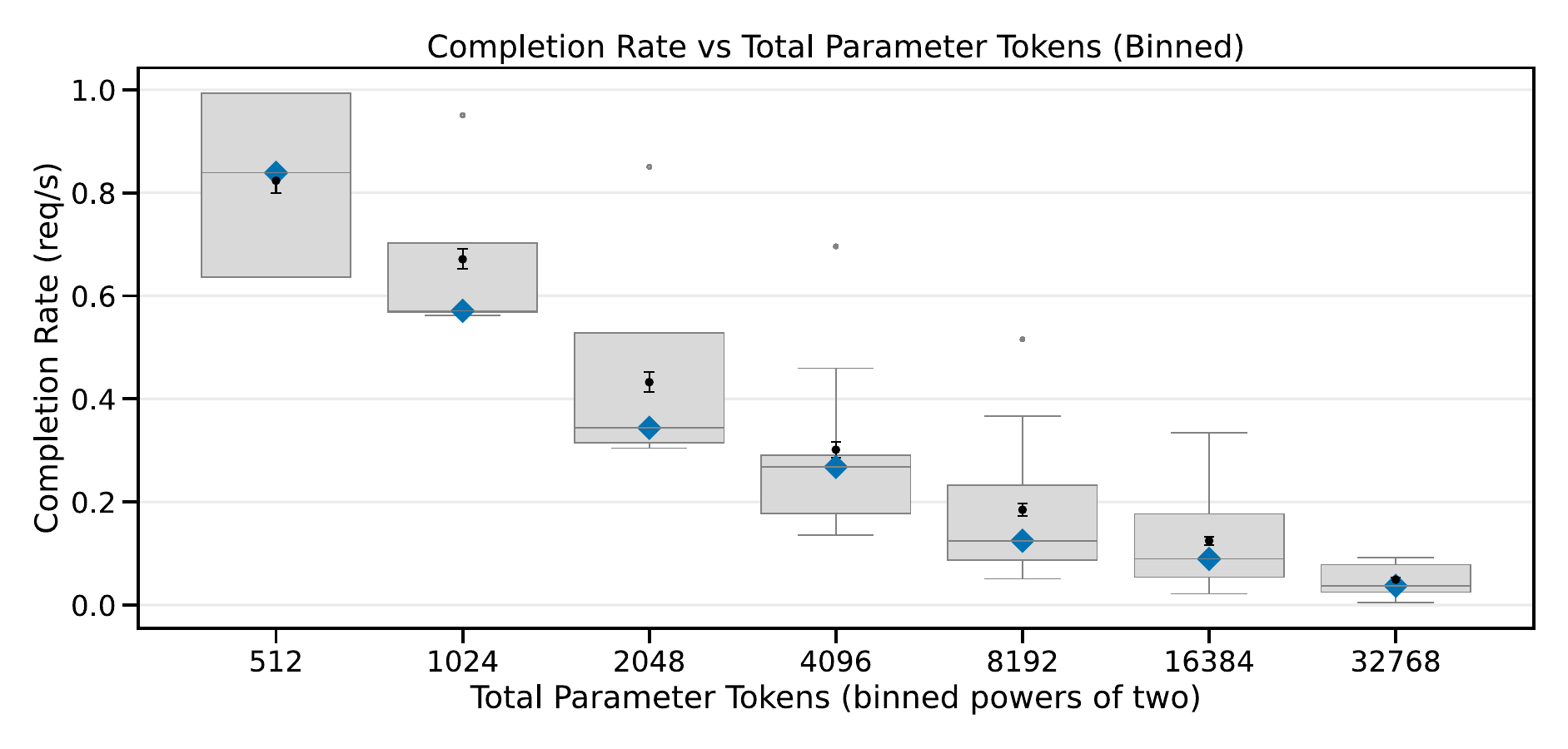}
        \caption{Completion Rate vs. Total\_Tokens (Binned)}
        \label{fig:completion_vs_tokens_binned}
    \end{subfigure}
    \hfill
    \begin{subfigure}{0.48\textwidth}
        \includegraphics[width=\linewidth]{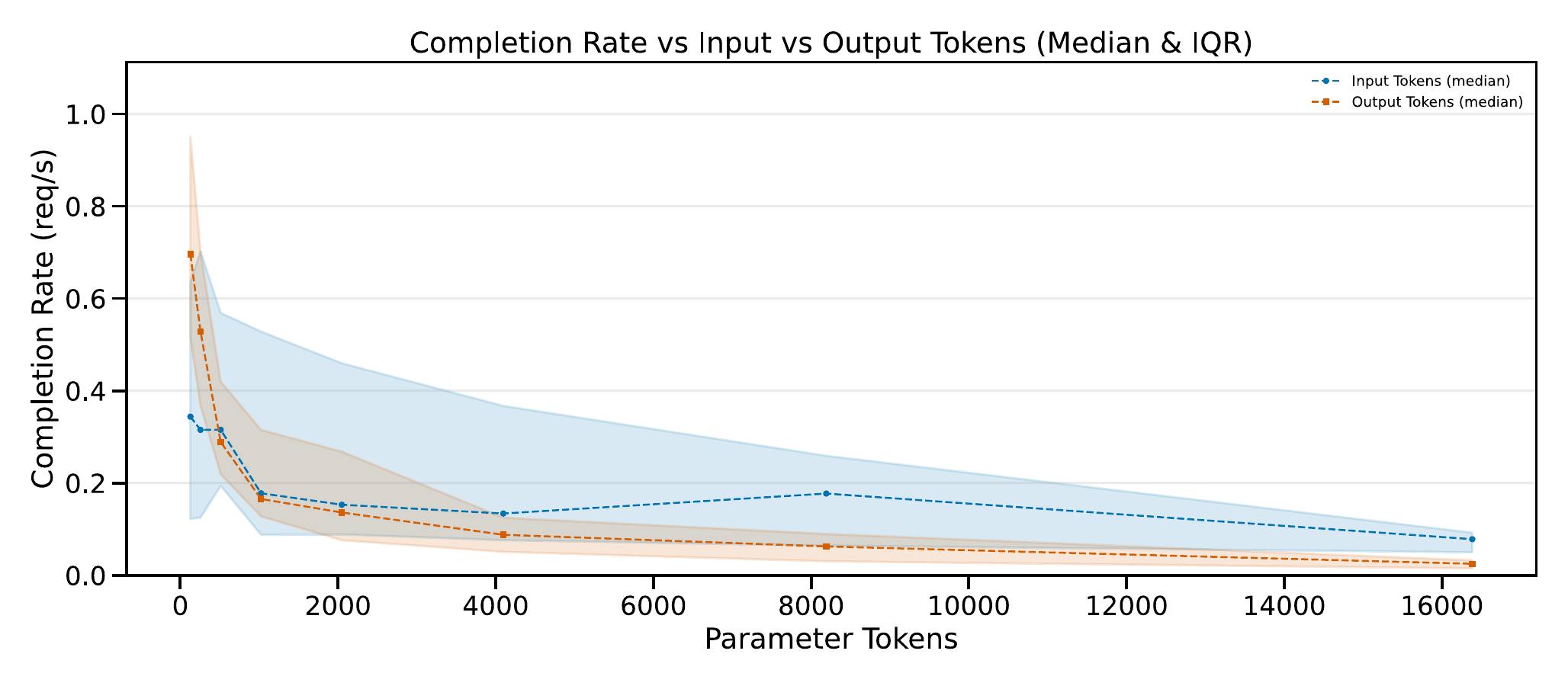}
        \caption{Completion Rate vs. Input/Output Tokens}
        \label{fig:completion_vs_tokens_line}
    \end{subfigure}
    \caption{System performance as a function of token count. Top row: End-to-end latency exhibits super-linear growth with Total\_Tokens, driven primarily by the number of output tokens. Bottom row: System completion rate (throughput) degrades sharply with token count, with input token length (prefill) emerging as the key bottleneck for overall throughput.}
    \label{fig:token_analysis}
\end{figure}

Our analysis reveals a critical trade-off between accommodating long sequences and maintaining system responsiveness, as detailed in Figure~\ref{fig:token_analysis}. End-to-end (E2E) latency scales super-linearly with the total token count, remaining modest up to 4096 tokens before accelerating sharply. This trend is accompanied by a significant increase in variance, indicating the onset of system contention and queuing effects. The breakdown between input and output tokens shows that latency is most sensitive to the number of generated tokens, pinpointing the autoregressive decoding phase as the primary driver of per-request computational cost.

Conversely, the system's overall throughput, measured by the completion rate \footnote{Completion rate measures how many requests per second are fully processed by the server, serving as a proxy for sustained capacity under load.}, shows a steep, monotonic decline with increasing token counts, dropping by over 20x from the 512-token to the 32768-token bin. While output token generation dictates the duration of individual requests, the analysis reveals that the length of the input prompt is the principal bottleneck for system throughput. The computationally intensive prefill \footnote{Prefill refers to the initial forward pass through the model for the input prompt before autoregressive decoding begins; it dominates runtime when prompts are long.} stage for long contexts appears to monopolize resources, severely limiting the rate at which new requests can be admitted and processed. This highlights a fundamental tension: while decode length drives user-perceived latency for a single request, it is the prefill cost of long prompts that saturates the system and throttles its multi-user capacity. Effective management of this trade-off is therefore essential, suggesting a need for strategies that optimize or mitigate the high cost of prefill for long-context requests.

\subsection*{Impact of concurrency on system efficiency}

\begin{figure}[htbp]
    \centering
    \begin{subfigure}{0.48\textwidth}
        \includegraphics[width=\linewidth]{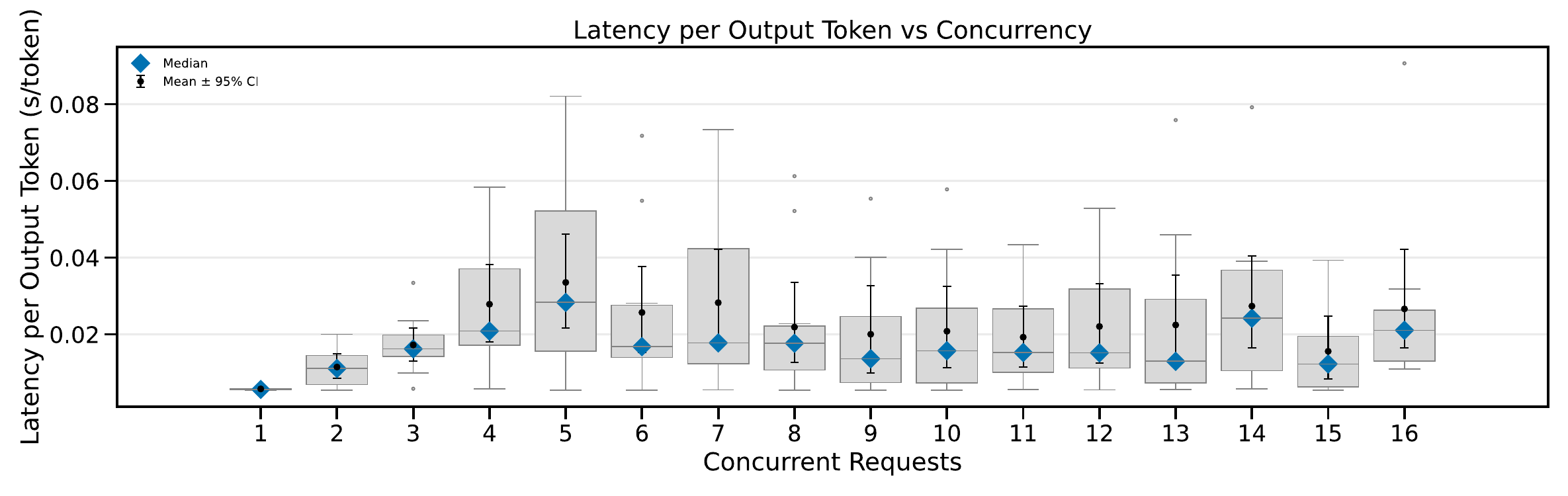}
        \caption{Latency per Output Token vs. Concurrency}
        \label{fig:latency_vs_concurrency}
    \end{subfigure}
    \hfill
    \begin{subfigure}{0.48\textwidth}
        \includegraphics[width=\linewidth]{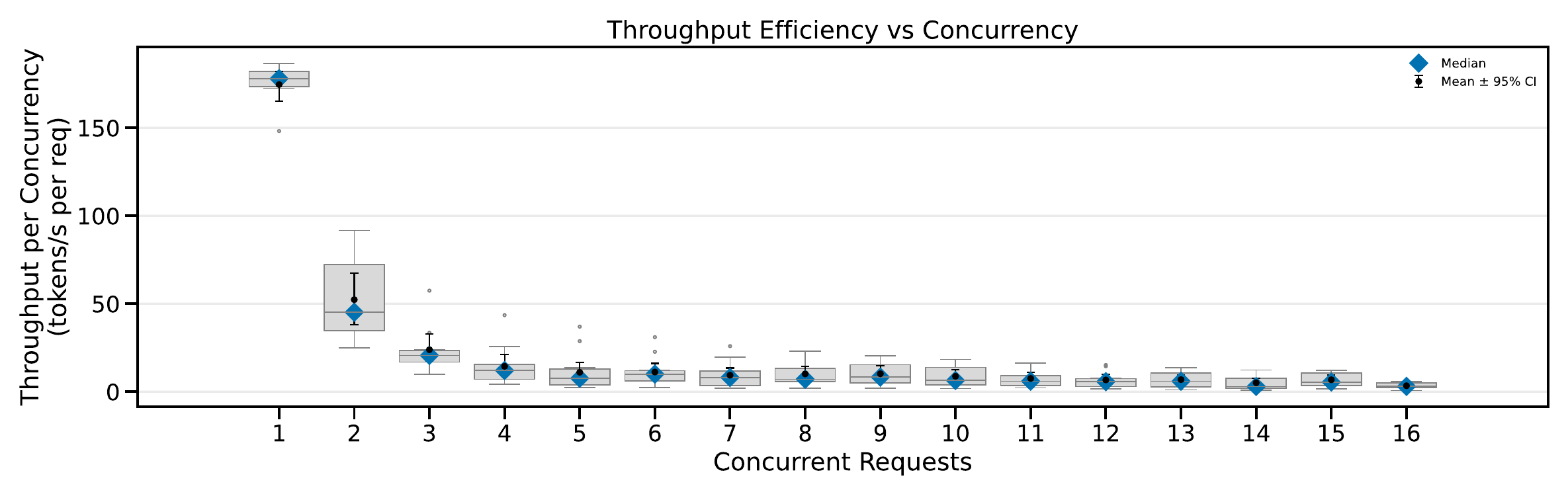}
        \caption{Throughput Efficiency vs. Concurrency}
        \label{fig:throughput_vs_concurrency}
    \end{subfigure}
    \caption{System performance under increasing concurrent loads}
    \label{fig:concurrency_analysis}
\end{figure}

The system's ability to scale with concurrent users is severely constrained by serialization overheads, as shown in figure above. The per-request throughput efficiency, a measure of productive work per user, collapses dramatically as the number of concurrent requests increases, falling by 75\% when moving from one to two users and declining further to a minimal plateau. This indicates that the system does not effectively parallelize the workload; instead, additional users appear to enter a queue, leading to substantial wait times rather than increased aggregate throughput.

This inefficiency is further reflected in the per-token latency. While latency initially spikes at low concurrency (peaking at 5 users), it stabilizes in a higher-latency regime (from 6 to 13 users) as dynamic batching \footnote{Dynamic batching combines multiple user requests into a single inference pass, improving GPU utilization but introducing variability in per-user latency.} begins to amortize some computational overhead. However, this stabilization comes at the cost of the catastrophic drop in per-request efficiency. The system finds a state of equilibrium where it processes batches at a steady, albeit slow, pace, but individual user experience is severely degraded. The presence of high-variance outliers across all concurrency levels further highlights intermittent contention and scheduling instability. These findings collectively suggest an operational sweet spot between 8-12 concurrent requests to balance utilization and predictability, but underscore that the serving strategy is fundamentally bottlenecked by its inability to perform true parallel processing. Achieving scalable performance will require advanced techniques like continuous batching and more sophisticated scheduling to overcome these serialization-induced limitations.

\paragraph{Comparative latency and throughput.}
Table~\ref{tab:cloud-vs-server} contrasts our single-GPU sovereign setup with representative cloud services, using reported average metrics (TTFT, throughput, and E2E latency). Cloud service numbers are drawn from independent artificialanalysis and openrouter services; server measurements are from our own evaluation.

\begin{table}[htbp]
\centering
\begin{tabular}{lcccc}
\toprule
\textbf{Model / Setup} & \textbf{Source} & \textbf{TTFT (s)} & \textbf{Throughput (t/s/user)} & \textbf{E2E latency (s)} \\
\midrule
\multirow{2}{*}{Gemini 2.5 Flash} 
    & artificialanalysis & 0.3  & 225   & 2.5 \\
    & openrouter         & 0.58 & 88.7  & —   \\
\midrule
\multirow{2}{*}{ChatGPT o3} 
    & artificialanalysis & 0.5   & 220   & 8.0 \\
    & openrouter         & 5.54  & 51.87 & —   \\
\midrule
\multirow{2}{*}{ChatGPT o1} 
    & artificialanalysis & 11.6  & 289   & 13.8 \\
    & openrouter        & 7.1   & 481   & —   \\
\midrule
\multirow{2}{*}{Claude 3.7 Sonnet} 
    & artificialanalysis & 14.5  & 66    & 17.1 \\
    & openrouter        & 1.52  & 53.74 & —   \\
\midrule
\multirow{2}{*}{DeepSeek R1} 
    & artificialanalysis & 103.5 & 20    & 128.6 \\
    & openrouter         & 0.75  & 59.63 & —   \\
\midrule
\multirow{2}{*}{\shortstack{Qwen3-30B\\(RTX 5090)}}
    & 1 user & 0.171 & 192.397    & 9.869 \\
    & 4 users         & 0.75  & 46.979 & 57.764   \\
\bottomrule
\end{tabular}
\newline
\caption{Comparison of TTFT, throughput, and estimated E2E latency across artificialanalysis.ai \citep{artificialanalysis2025} and openrouter.ai \cite{openrouter2025} benchmarks.}
\label{tab:cloud-vs-server}
\end{table}

\section{Discussion}

\label{sec:discussion}
\paragraph{Key findings.} 
Our results show that a sovereign deployment of \texttt{Qwen3-30B-A3B} on a single RTX~5090 can provide competitive model quality and near-cloud interactivity for individual users. Time-to-first-token is faster than many cloud baselines, and throughput remains high under light load. However, concurrency quickly degrades efficiency: per-request throughput drops steeply beyond two users, and end-to-end latency rises super-linearly with prompt and output length. These results indicate that current consumer hardware is well-suited for private single-user or small-team workloads, but multi-user scalability remains a bottleneck.

\paragraph{Implications.} 
The observed bottlenecks arise primarily from prefill costs, limited host–GPU coordination, and serialization effects in the serving stack. Addressing these constraints will require runtime-level innovations (e.g., paged attention, dynamic batching, speculative decoding) and potentially multi-GPU scaling. For practitioners, this implies that sovereign deployments in 2025 are viable for latency-sensitive but lightly concurrent use cases, while production-scale services will demand more advanced scheduling and hardware parallelism.

\paragraph{Limitations.} \label{sec:limitations}
Several factors constrain the generality of our findings. First, synthetic workloads were generated using \texttt{llmperf}, which approximates but does not fully capture the diversity of real conversational traffic. Second, we evaluated only one model and one GPU, limiting cross-hardware and cross-model comparability. Third, all measurements were obtained in a localhost setting, excluding the impact of network latency and production orchestration. Finally, we relied solely on \texttt{llama.cpp}; alternative frameworks such as vLLM may significantly shift throughput and tail-latency trade-offs. Future work should expand the hardware–software design space, incorporate real usage traces, and explore runtime optimizations to produce robust cost–QoE curves for sovereign deployments.

\section{Conclusion}

This study demonstrates that a single RTX 5090 running a quantized \texttt{Qwen3-30B-A3B} can deliver high-quality, private inference with latency and throughput comparable to leading cloud services, provided workloads remain light and concurrency low. For multi-user scenarios, however, system efficiency collapses due to prefill overheads, serialization, and limited host–GPU coordination. These findings establish the minimal hardware regime for sovereign deployments in 2025 while underscoring that scalable, cloud-parity performance will require advances in runtimes, scheduling, and multi-GPU parallelism.

It is important to note, however, that the observed throughput drop is most pronounced under perfectly simultaneous request bursts. In practice, user queries typically arrive with some degree of temporal staggering, which naturally mitigates queueing contention. As a result, the degradation in responsiveness is often less severe than suggested by worst-case stress tests, and end users may not perceive the full extent of the concurrency limitations in real-world usage.

\section*{Acknowledgements}
We would like to thank Lepage Théo for his guidance before the start of this work, advising us on how to structure and plan the article, as well as for his final review and feedback, which provided valuable insights.

\newpage

\bibliographystyle{plainnat} 
\bibliography{references}    

@inproceedings{frantar2022gptq,
  title     = {OPTQ: Accurate Post-Training Quantization for Generative Pre-trained Transformers},
  author    = {Frantar, Elias and Ashkboos, Saeed and Gholami, Amir and Alistarh, Dan},
  booktitle = {International Conference on Learning Representations (ICLR)},
  year      = {2023},
  url       = {https://openreview.net/forum?id=tcbBPnfwxS}
}

@article{lin2024awq,
  title   = {AWQ: Activation-Aware Weight Quantization for LLM Compression and Acceleration},
  author  = {Lin, Jiayi and Tang, Sheng and Li, Xu and Wang, Hongyi and He, Yihang and Li, Mu and Chen, Zhiqiang and Wang, Yizhou},
  journal = {arXiv preprint arXiv:2306.00978},
  year    = {2024}
}

@article{dettmers2023spqr,
  title   = {SpQR: A Sparse-Quantized Representation for Near-Lossless LLM Weight Compression},
  author  = {Dettmers, Tim and Lewis, Mike and Shleifer, Sam and Zettlemoyer, Luke},
  journal = {arXiv preprint arXiv:2306.03078},
  year    = {2023}
}

@article{xia2024fp6,
  title   = {FP6-LLM: Efficiently Serving Large Language Models Through FP6-Centric Algorithm-System Co-Design},
  author  = {Xia, Haojun and Zheng, Zhen and Wu, Xiaoxia and Chen, Shiyang and Yao, Zhewei and Youn, Stephen and Bakhtiari, Arash and Wyatt, Michael and Zhuang, Donglin and Zhou, Zhongzhu and Ruwase, Olatunji and He, Yuxiong and Song, Shuaiwen Leon},
  journal = {arXiv preprint arXiv:2401.14112},
  year    = {2024}
}

@misc{llamacpp2025,
  title        = {llama.cpp: Inference of LLaMA and other LLMs in C/C++},
  author       = {Georgi Gerganov and contributors},
  year         = {2025},
  howpublished = {\url{https://github.com/ggerganov/llama.cpp}}
}

@article{paloniemi2025onpremise,
  title   = {On-Premise Large Language Model Deployments: Motivations, Challenges, and Case Studies},
  author  = {Paloniemi, Tuomas and Nieminen, Antti and Rossi, Pekka},
  journal = {Journal of Cloud Computing},
  year    = {2025}
}

@misc{sovereignAI2024,
  title        = {Sovereign AI: Policy and Infrastructure Strategies for National LLM Hosting},
  author       = {{European Commission AI Office}},
  year         = {2024},
  howpublished = {\url{https://digital-strategy.ec.europa.eu}}
}

@article{wang2023ondevicellms,
  title   = {On-Device Language Models: A Comprehensive Review},
  author  = {Xu, Jiajun and Li, Zhiyuan and Chen, Wei and Wang, Qun and Gao, Xin and Cai, Qi and Ling, Ziyuan},
  journal = {arXiv preprint arXiv:2409.00088 },
  year    = {2024}
}

@misc{qwen32025release,
  title        = {Qwen3 Technical Report and Model Release},
  author       = {Alibaba DAMO Academy},
  year         = {2025},
  howpublished = {\url{https://huggingface.co/Qwen}}
}

@inproceedings{kwon2023pagedattention,
  title     = {Efficient Memory Management for Large Language Model Serving with PagedAttention},
  author    = {Kwon, Woojeong and Lin, Yizhuo and Xie, Xuechen and Chen, Tianqi},
  booktitle = {Proceedings of the 29th ACM Symposium on Operating Systems Principles (SOSP)},
  year      = {2023}
}

@inproceedings{zhang2023fastserve,
  title     = {FastServe: Efficient LLM Serving Using Speculative Scheduling},
  author    = {Zhang, Zhihao and Xu, Hang and Wang, Yuxin and Chen, Kai},
  booktitle = {Proceedings of the ACM Symposium on Cloud Computing (SoCC)},
  year      = {2023}
}

@article{chitty2024llminferencebench,
  title={LLM-Inference-Bench: Inference Benchmarking of Large Language Models on AI Accelerators}, 
      author={Krishna Teja Chitty-Venkata and Siddhisanket Raskar and Bharat Kale and Farah Ferdaus and Aditya Tanikanti and Ken Raffenetti and Valerie Taylor and Murali Emani and Venkatram Vishwanath},
      year={2024},
  journal = {arXiv preprint arXiv:2411.00136}
}

@inproceedings{dao2022flashattention,
  title     = {FlashAttention: Fast and Memory-Efficient Exact Attention with IO-Awareness},
  author    = {Dao, Tri and Fu, Daniel and Ermon, Stefano and Rudra, Atri and Re, Christopher},
  booktitle = {Advances in Neural Information Processing Systems (NeurIPS)},
  year      = {2022}
}

@article{shen2024flashattention2,
  title   = {FlashAttention-2: Faster Attention with Better Parallelism and Work Partitioning},
  author  = {Shen, Haotian and Dao, Tri and Chen, Zhewei and Song, Xinyun and Zhao, Tianle and Li, Zhuohan and Stoica, Ion and Gonzalez, Joseph E. and Zaharia, Matei},
  journal = {arXiv preprint arXiv:2307.08691},
  year    = {2024}
}

@inproceedings{shazeer2021switch,
  title = {Switch Transformers: Scaling to Trillion-Parameter Models with Simple Routing},
  author = {Shazeer, Noam and others},
  booktitle = {Proceedings of the 9th International Conference on Learning Representations (ICLR)},
  year = {2021},
  note = {arXiv:2101.03961},
  url = {https://arxiv.org/abs/2101.03961}
}

@misc{vllm2023,
  title = {{vLLM}: A fast and memory-efficient LLM serving library (repo \& docs)},
  author = {{vLLM Project}},
  year = {2023},
  howpublished = {\url{https://github.com/vllm-project/vllm}},
  note = {PagedAttention and vLLM documentation}
}

@misc{llmperf2024,
  title = {LLMPerf: A benchmarking and load-generation tool for LLM inference},
  author = {{LLMPerf Project}},
  year = {2024},
  howpublished = {\url{https://github.com/ray-project/llmperf}},
  note = {Tool used to generate synthetic LLM workloads in experiments}
}

@misc{lmeval2023,
  title = {lm-evaluation-harness: A framework for evaluating language models},
  author = {Ethayarajh, K. and contributors (EleutherAI)},
  year = {2023},
  howpublished = {\url{https://github.com/EleutherAI/lm-evaluation-harness}},
  note = {Evaluation harness used for MMLU and comparable benchmarks}
}

@misc{aime2024_hf,
  title = {AIME 2024 Dataset},
  author = {Maxwell-Jia and HuggingFace Datasets contributors},
  year = {2024},
  howpublished = {\url{https://huggingface.co/datasets/Maxwell-Jia/AIME_2024}},
  note = {Dataset of AIME 2024 problems (used for math reasoning evaluation)}
}

@misc{artificialanalysis2025,
  title = {Artificial Analysis — AI Model \& API Providers Analysis},
  author = {{ArtificialAnalysis}},
  year = {2025},
  howpublished = {\url{https://artificialanalysis.ai}},
  note = {Model comparison and independent benchmarks (source for Table 1 numbers)}
}

@misc{openrouter2025,
  title = {OpenRouter — Models \& Provider Metrics},
  author = {{OpenRouter}},
  year = {2025},
  howpublished = {\url{https://openrouter.ai}},
  note = {Aggregated provider and model metrics (source for Table 1 numbers)}
}

@article{mckay1979lhs,
  title     = {A Comparison of Three Methods for Selecting Values of Input Variables in the Analysis of Output from a Computer Code},
  author    = {McKay, Michael D. and Beckman, Richard J. and Conover, William J.},
  journal   = {Technometrics},
  volume    = {21},
  number    = {2},
  pages     = {239--245},
  year      = {1979},
  url       = {https://doi.org/10.1080/00401706.1979.10489755}
}

\end{document}